# Superconducting joint between iron-based superconductor tapes

Yanchang Zhu[1,3], Dongliang Wang[1,2,*], Chundong Zhu[3], He Huang[1,2], Zhongtang Xu[1,2], Shifa Liu[1,2], Zhe Cheng[1,2], and Yanwei Ma[1,2,*]

[1] Key Laboratory of Applied Superconductivity, Institute of Electrical Engineering, Chinese Academy of Sciences, Beijing 100190, People's Republic of China

[2] University of Chinese Academy of Sciences, Beijing 100049, People's Republic of China

[3] School of Materials Science and Engineering, Wuhan University of Technology, Wuhan 430070, China

**Abstract**

Superconducting joints are essential for iron-based superconductor's applications in future. In this study, a process for fabricating superconducting joints between $Sr_{1-x}K_xFe_2As_2$ (Sr-122) tapes is developed for the first time. The Ag sheath was peeled off from one side of each sample. The exposed superconducting parts of the two tapes were joined and wrapped again with Ag foil. The diffusion bonding of the iron-based superconducting joint was achieved by hot-pressing process in Argon atmosphere. The superconducting properties, microstructures and the elements distribution of the joint regions had been investigated. The pressure and pressing times were optimized in order to enhance the transport current of the joints. At 4.2 K and 10 T, a transport critical current $I_c$ of 57 A for the joint was obtained, which is approximately 63.3% of the current capacity of the tapes themselves. Furthermore, the joint resistances d$V$/d$I$ were estimated from the $V$-$I$ curve of the joints and the calculated joint resistances values are below $10^{-9}$ Ω. These results demonstrate that the hot pressing was useful for fabricating the superconducting joint samples.

**Keywords:** superconducting joint, iron-based superconductor, transport property, diffusion bonding, microstructures

[*]Author to whom correspondence should be addressed: E-mail: dongliangwang@mail.iee.ac.cn or ywma@mail.iee.ac.cn

## 1. Introduction

Iron-based superconductors have become a major topic of discussion these years due to their unique performance, such as a high superconducting transition temperatures ($T_c$) up to 55 K [1, 2], large transport $J_c$ over 1 MA/cm$^2$ achieved in thin films [3-5], and high upper critical field above 100 T [6, 7] with low anisotropy (< 2 for Sr/BaKFeAs) [8, 9]. In the past years, significant progresses towards high-performance iron-based conductors were made [10-13], the $J_c$ value of Ba-122 wire and tapes had been rapidly improved to $1.5 \times 10^5$ A/cm$^2$ at 4.2 K and 10 T in short samples [14], and the 100 m class Sr$_{1-x}$K$_x$Fe$_2$As$_2$ (Sr-122) tapes were fabricated last year [15]. However, for large-scale use, it is very common to connect several pieces of long iron-based superconducting tapes. Hence, superconducting joint is very important for the merits of lowering the total heating generation and persistent current operation. In order to operate the magnet in persistent mode, the joints between the coils must be superconducting even under the high magnetic field. Therefore, joining between two superconducting wires or tapes has become one of the most important technologies.

So far, techniques of joining superconductor wires/tapes such as NbTi, BiSCCO and MgB$_2$ wires/tapes have been well developed [16-25]. For the low temperature superconductors, especially for NbTi, jointing methods are very mature and convenient in the manufacture of persistent mode (PM) magnets. Soldering with solder alloy is one of the choices in commercial magnet manufacture, and the critical current of the joints can reach as high as 1000 A at 4.2 K and 1 T [20], with resistance in the range of 10$^{-13}$~10$^{-14}$ Ω. Cold press is another good choice to connect the NbTi filaments [21, 22].

The current of joints fabricated by this method was as high as 468 A (1 T, 4.2 K). Persistent $MgB_2$ joints have been made predominantly by powder in tube (PIT) techniques which are similar to those used in wire production. The critical current ratio (CCR) value can reaches 90% at 4.2 K and joint resistance reaches $10^{-10} \sim 10^{-12}$ Ω [17]. Oomen [23] used the hot-pressing technique at slightly lower temperatures between 600 and 700 ℃ to fabricate joints with dense and well-connected filler materials. CCR values for these joints are only 25~50% at 4.2 K and 1 T, and the resistance below $10^{-9}$ Ω. The formation of joints between BiSCCO is typically fabricated by repeating the heat treatment process used for the conductor itself. And this involves cold pressing of the exposed filaments of particularly reacted tapes followed by the same extensive thermomechanical treatment used to form the BiSCCO phase in the conductor [24, 25]. The critical current of the joints could be higher than the tapes as reported by Peng Chen [16], and the joint resistance was estimated to be below $10^{-12}$ Ω at 4.2 K and self-field. In addition, a ground-breaking demonstration has been made in the joint methods of the REBCO coated conductors [26, 27]. Park [26] prepared a well bonded lap joint by partial melting of the REBCO layers under low $PO_2$. The critical current (84 A) of the joint at 77 K and self-field was achieved, and joint resistance less than $10^{-17}$ Ω was confirmed by using a decay measurement at 77 K and self-field [26].

Until now, joining iron-based superconducting tapes hasn't been reported and the joining method is much more difficult. At first, as the same as the condition of the cuprate superconductors, the iron-based superconducting core are fragile and harder than the sheath materials. Thus, it is very difficult to join two pieces of iron-based superconductor composite tape simply by cold-pressing, like NbTi's case. On the other hand, the iron-based superconductor is more sensitive with the air, compared with the cuprate superconductors, which means that the sintering process should better be made in a protective atmosphere. In general, the lap joint method and diffusion bonding which shows good joint properties of joint between the cuprate superconductors can be adopted in this study. Besides, it was found that the hot pressing (HP) progress could significantly increases the transport $J_c$ of Sr-122 tape by improving core density and grain alignment, simultaneously [12]. Recently, the transport $J_c$ value up to 0.15 MA/cm$^2$ at 4.2 k and 10 T has been obtained in Ba-122 tapes through HP method [14]. According to these results, it is possible to form Sr-122 joints between two pieces of Sr-122 tapes by using HP method, during this process, the Sr-122 grains of the connecting area could heal better and connect them together.

In this paper, a method of joining Sr-122 tapes using HP was first proposed. Samples of single filament Sr-122 tapes were joined by this progress. The critical current $I_c$ and the joint resistance of samples were measured at 4.2 K and in different magnetic fields. The $I_c$ was 57 A at 4.2 K and 10 T, meanwhile the joint resistance estimated from the *V-I* curve was below $10^{-9}$ Ω.

**2. Experimental details**

The Sr-122 tapes were fabricated by the *ex-situ* PIT method. Firstly, Sr fillings, K pieces, Fe and As powders were mixed by ball-milling method. Then the mixed powders were loaded into Nb tubes and then sintered at 900 °C for 35 h. The sintered superconducting bulk was ground into powders under Ar atmosphere, and then the Sn material was added to increase grain connectivity. The fine powders were finally packed into Ag tubes with OD 8 mm and ID 5 mm. These tubes with mono-core inside were finally deformed into tapes with thickness of 0.4 mm. In the end, short samples were cut from the long tapes, and then sintered at 880 °C for 0.5 h in Argon atmosphere.

Figure 1 shows a schematic view of the method of assembling. The sides of the sheath materials were partially peeled off mechanically. The width of the joints was 3.5 mm, and the wrap length was 6 mm. None of bonding agent was used. The joint samples were heat treated in Argon atmosphere and the thermal circle is the same with the heat treatment process of the non-joint tapes. We did a series of controlled experiments to investigate the effect of pressure and pressing time on the joint sample properties. The parameters of the hot-pressing process were shown in table 1.

We used the X-ray Computed Tomography (XCT, Model: GE Microme|x) to analyze the cracks distribution of the connection area of the joint samples. Microstructures in joint area were observed by Electron Probe Microanalysis (EPMA, Model: JEOL JXA8230). The temperature dependence of the resistivity at 0 T were carried out using a four-probe method by PPMS (Model: PPMS-9). Transport critical currents ($I_c$) of the samples were measured at 4.2 K using a standard four-probe technique, with a criterion of 1 $\mu$V cm$^{-1}$. The magnetic field dependence of transport $I_c$ values for all samples were evaluated at the High Field Laboratory for Superconducting Materials (HFLSM) in Sendai or Institute of Plasma Physics, CAS in Hefei.

## 3. Results and discussions

The SEM images of longitudinal cross section of the joint samples under different pressure values are shown in figure 2. It reveals that there are a lot of micro-cracks in the whole cross section from figure 2(a). In figure 2(b) the micro-cracks in the tape area decrease and are mainly concentrated in the joint region. When the pressure further increases to 13.8 MPa there are almost no obvious micro-cracks in the tape area. Meanwhile, as the pressure increased, the number of micro-cracks around the connection area was also getting less. It indicates that the micro-cracks in the whole joint section significantly decreased with increasing the pressure. This means that as the pressure increases, the tapes are well assembled, which may load large transport current.

As shown in figure 3, the temperature dependence of the resistance at self-field for all the samples was measured using a typical standard four-probe method at 4.2 K and self-field. Table 2 lists the $T_c^{onset}$, $T_c^{zero}$ and transition width of all the samples. We can see from the curves that the samples HP300 and HP200 shows sharp transition compared to the HP50 and HP100. The $T_c^{zero}$ values of the joint samples are increased as the pressure increases, indicating that the impurity phase formed in the connection area significantly affect the superconducting transition of the joint samples. Furthermore, the smaller transition width $\Delta T_c$ of 1.2 K and 1 K obtained for the samples HP200 and HP300 which are very close to that of the tapes; while the ones of HP50 and HP100 are 2.4 K and 2.3 K, respectively. These results suggest that the superconducting phases of the HP300 and HP200 are more homogeneous.

In order to further characterize the properties of the samples, we tested the critical current of all samples. The voltage and current terminals were attached on either side of the joint to measure voltage of the joint sample. The distance between the voltage terminals is 20 mm for all samples. As an example, the voltage of HP200 at 4.2 K and 10 T was plotted in figure 4(a). The calculated transport $I_c$ of HP200 was 40 A at 4.2 K and 10 T, with a criterion of 1 $\mu$V cm$^{-1}$. The joint resistances d$V$/d$I$ estimated from the $V$-$I$ curve were below $10^{-9}$ Ω. But the accuracy of this $V$-$I$ measurement system is $10^{-9}$ Ω, thus the accurate resistance data should be measured by using the inductive method, which was not mentioned in this work. Figure 4(b) presents the magnetic-field dependence of transport $I_c$ at 4.2 K for the Sr-122 joint tapes hot-pressed at different pressure. The applied fields up to 10 T were parallel to the joint sample surface. The critical current of sample HP50 is too low to be measured in the magnetic field over 2 T. As is evident from the figure, the $I_c$ increases monotonically with the increases of HP pressure load up to 9.2 MPa, and then rapidly decreased when further increasing pressure load to 13.8 MPa. The best Sr-122 joint samples HP200 exhibit a critical current ratio (CCR=$I_c^{joint}$/$I_c^{non-joint}$) of 35.3% at 10 T and 4.2 K, which is firstly reported so far for iron-based tapes. Besides, the superconducting current as high as 178 A was also achieved at 4.2 K and self-field. And the CCR of samples HP100 and HP300 are 11.3% and 12.8% at 4.2 K and 10 T, respectively. It can be inferred that the pressure is beneficial for the homogeneous superconducting phase and healing of longitudinal cracks in the connection area. However, the transport performance of sample HP300 is not the best, which means that the microstructure of joint region should be further analyzed.

Figure 5 shows the X-ray Computed Tomography (XCT) images of joint samples

HP50、HP100、HP200 and HP300, respectively. As shown in figure 5, all the samples have longitudinal macro-cracks along the direction of current transport, and the longitudinal macro-cracks decrease with increasing the pressure. At the same time, the transverse macro-cracks occur when the pressure increased to 13.8 MPa. These results indubitably indicate that all the cracks are harmful to the transport performance of the joints, but the effect of transverse macro-cracks on blocking the current is larger than that of longitudinal macro-cracks. It may be the main reason why the critical currents of joint samples were not improved continuously with increasing the pressure. Although the sample HP300 shows more homogeneous superconducting phase and less longitudinal cracks in the connection area than the other samples, the CCR of sample HP200 is larger than that of HP300, when the applied pressure is 9.2 MPa.

To characterize the performance of joints more adequately, EPMA mapping images are shown in figure 6 to further demonstrate the element distribution of sample HP300. The elements Sr, K, Fe and As of superconducting phase are almost homogeneously distributed in the tape area. However, the element K is seriously deficient around the connection area, especially along the crack, which can also explain the low superconducting performance of the joint, compared to the non-joint tapes. This also provides us with some ideas for further improving the performance of the joint, which is eliminating cracks and preventing K loss.

The CCR of the joint was not high and there were some limitations to the properties of the joints. Our results show that increasing the pressure can eliminate longitudinal cracks, however, the transverse cracks will appear. Meanwhile, the K loss is considered to be happened during the high temperature heat-treatment without pressure. Therefore, we tried to reduce the pressure and increase the pressing time, in order to further enhance the transport current of the joints. Figure 7 shows the magnetic field dependence of transport critical current at 4.2 K for joint tapes, which were hot pressed with low pressure and long pressing time. We can see that the best joint sample was obtained when the pressing time is 4.5 h and the pressure is 2.76 MPa. The critical current of sample CHP60 was 57 A and the CCR of the joint is 63.3% at 4.2 K and 10 T, which has been greatly improved compared to the sample HP200. Even for the joint HP50, when the pressing time increased from 30 min to 4.5 h, the $I_c$ of 57 A and the CCR of 63.3% at 4.2 K and 10 T were achieved for sample CHP60. However, when the pressure enhanced into 3.22 MPa, the transport current of joint decreased dramatically.

Figure 8 shows the SEM and EPMA mapping images of longitudinal cross section for the joint sample CHP60. It is obvious that the cracks almost disappeared in tape area, which is different from those of HP50 and HP100. Besides, as presented in Figure 8, the loss element potassium of CHP60 is smaller than that of HP300. Therefore, it can be deduced that the low pressure with a long pressing time is more beneficial to avoid the elements losing and cracks around the connection area.

Figure 9 shows the XCT images of joint samples CHP30, CHP40, CHP60 and CHP70, respectively. As shown in figure 9, these joint samples have both longitudinal and transverse macro-cracks in the joint area. For all material except CHP70, longitudinal cracks decrease as pressure increases while the transverse cracks increase as pressure increases, which is as same as the result of figure 5. When the pressing time was 0.5 h, the best transport property was achieved for the joint HP200 pressed under 9.2 MPa, in which there were no obvious transverse cracks could be seen. Once the pressure was increased from 9.2 MPa to 13.8 MPa, the significant transverse cracks of the joint could be observed, thus the $I_c$ of the joint HP300 was decreased dramatically. When the pressing time was 4.5 h, the best transport property was achieved for the joint CHP60 pressed under 2.76 MPa, in which there were a few of transverse cracks could be seen. Once the pressure was increased from 2.76 MPa to 3.22 MPa, more transverse cracks appeared, then the transport current of the joint CHP70 was decreased. Therefore, there may be an "optimal" point which contains a certain ratio of these two cracks that offer the best property for the joints. However, for different pressing time, the "optimal" point was also different.

According to the results shown above, it is clear that the transport properties of iron-based superconducting joints are affected mainly by the superconducting phases and microstructures in the connection area. Firstly, in this work, three kinds of cracks were observed: longitudinal macro-cracks, transverse macro-cracks and micro-cracks. The macro-cracks are mainly in the tapes which were connected; while the micro-cracks are mainly distributed between the tapes which were connected. Transverse macro-cracks are more harmful to the transport performances of tapes, compared to the longitudinal ones. Meanwhile, the micro-cracks will block the current transport from one tape to another tape for the joint. Therefore, theses cracks are should be avoided in the joint section. Secondly, the element K loss in the joint area should also be avoided. Longer pressing time can obviously prevent the element K from losing, compared with the shorter pressing time. That's because the loss of potassium element was happened at high temperature, thus extending pressing time can eliminate this problem. Thirdly, the texture degree in the connection area is also very important, though it was not mentioned in this work. It is well-known that $c$-axis texture should be induced to relieve the weak-link effect at grain boundaries for iron-based superconducting samples. However, in our case, the crack problem should be resolved firstly before achieving high texture degree in the connection area. The bonding agent could be useful for eliminating the cracks in joint samples. In the future, further increase of the transport current for joint sample is expected by tailoring diffusion bonding parameters and bonding agent.

## 4. Conclusion

In summary, we successfully fabricated the world's first superconducting joint of iron-based Sr-122 tapes. The joint samples were hot pressed under various pressures of 1.38 MPa~13.8 MPa and pressing time of 0.5 h or 4.5 h. It was found that the pressure and pressing time significantly affect the CCR of the joint samples. Our results show that the longer pressing time is benefit for reducing the K loss. Meanwhile, the lower and higher pressures are both not suitable to assembling two superconducting cores. The highest CCR of 63.3% (at 4.2 K in 10 T) is obtained for the Sr-122 joint sample pressed at 2.76 MPa with pressing time of 4.5 h.

**Acknowledgments**

The authors would like to thank Prof. S. Awaji at Sendai, and Dr. Fang Liu, Mr. Lei Lei, Prof. Huajun Liu at Hefei for $I_c$-$B$ measurement. This work is partially supported by the National Natural Science Foundation of China (Grant Nos. 51577182 and 51320105015), the Beijing Municipal Science and Technology Commission (Grant No. Z171100002017006), the Bureau of Frontier Sciences and Education, Chinese Academy of Sciences (QYZDJ-SSW-JSC026), the Key Research Program of the Chinese Academy of Sciences (Grant No. XDPB01-02).

**Table 1** Parameters of hot-pressing process

| Samples | HP50 | HP100 | HP200 | HP300 | CHP30 | CHP40 | CHP50 | CHP60 | CHP70 |
|---|---|---|---|---|---|---|---|---|---|
| Pressure (MPa) | 2.3 | 4.6 | 9.2 | 13.8 | 1.38 | 1.84 | 2.3 | 2.76 | 3.22 |
| Pressing time (h) | 0.5 | 0.5 | 0.5 | 0.5 | 4.5 | 4.5 | 4.5 | 4.5 | 4.5 |

**Table 2.** Transition width of the joint samples and tape

| Samples | $T_c^{onset}$ (K) | $T_c^{zero}$ (K) | $\Delta T_c$(k) |
|---------|-------------------|------------------|-----------------|
| HP50    | 32.7              | 30.39            | 2.4             |
| HP100   | 33.99             | 31.69            | 2.3             |
| HP200   | 34.65             | 33.45            | 1.2             |
| HP300   | 35.5              | 34.5             | 1               |
| Tape    | 34.57             | 33.84            | 0.7             |

**Captions**

Figure 1. A schematic picture of joining process: (a) peeling off the sheath metal; (b) wrapping in Ag foil.

Figure 2. SEM images of longitudinal cross section of Sr-122 superconducting joints under 0.5 h pressing time.

Figure 3. Temperature dependence of resistivity for the non-joint and joint tapes under 0.5 h pressing time.

Figure 4. (a) *I-V* characteristics of the joint part for HP200 measured at 4.2 K and 10 T. (b) Magnetic field dependence of transport critical current at 4.2 K for non-joint and joint samples under 0.5 h pressing time.

Figure 5. X-ray Computed Tomography images of the joint samples under 0.5 h pressing time.

Figure 6. SEM and EPMA mapping images of the joint sample HP300.

Figure 7. Magnetic field dependence of transport critical current at 4.2 K for non-joint and joint tapes under 4.5 h pressing time.

Figure 8. SEM and EPMA mapping images of the joint sample CHP60.

Figure 9. XCT images of the joint samples under 4.5 h pressing time.

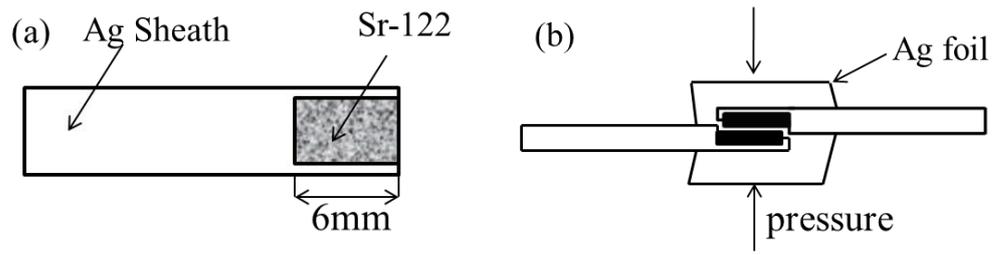

Figure 1. Zhu et al

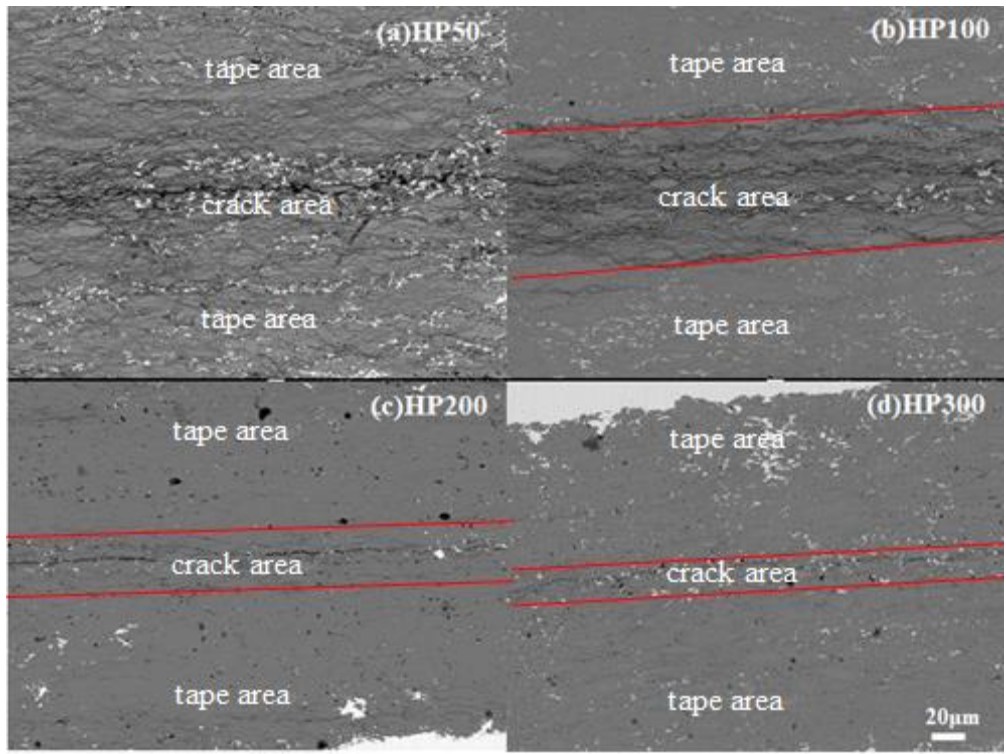

Figure 2. Zhu et al

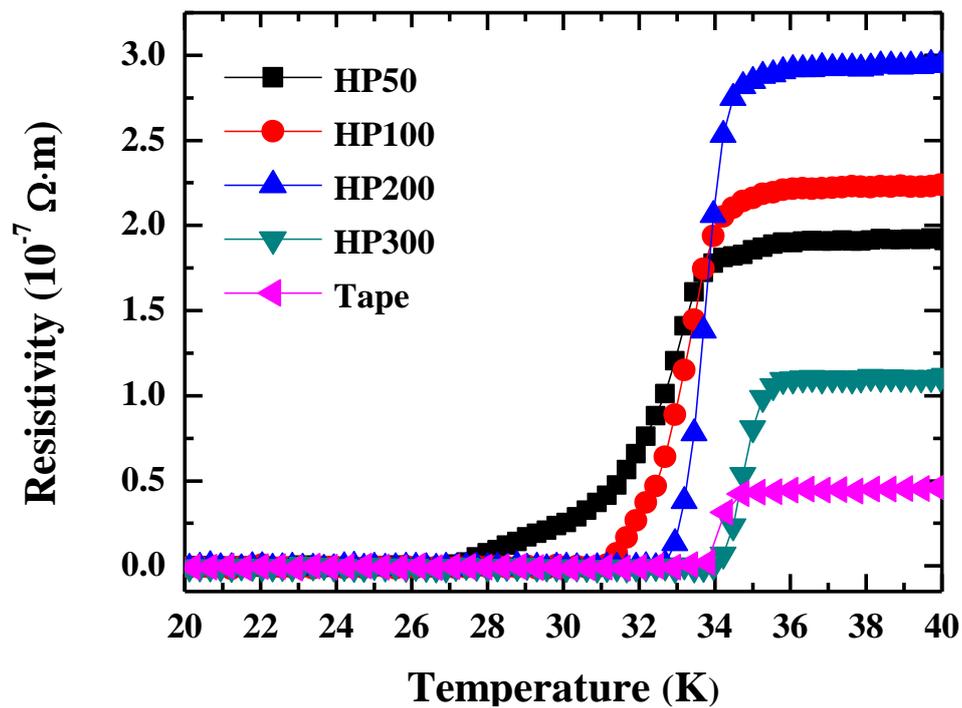

Figure 3. Zhu et al

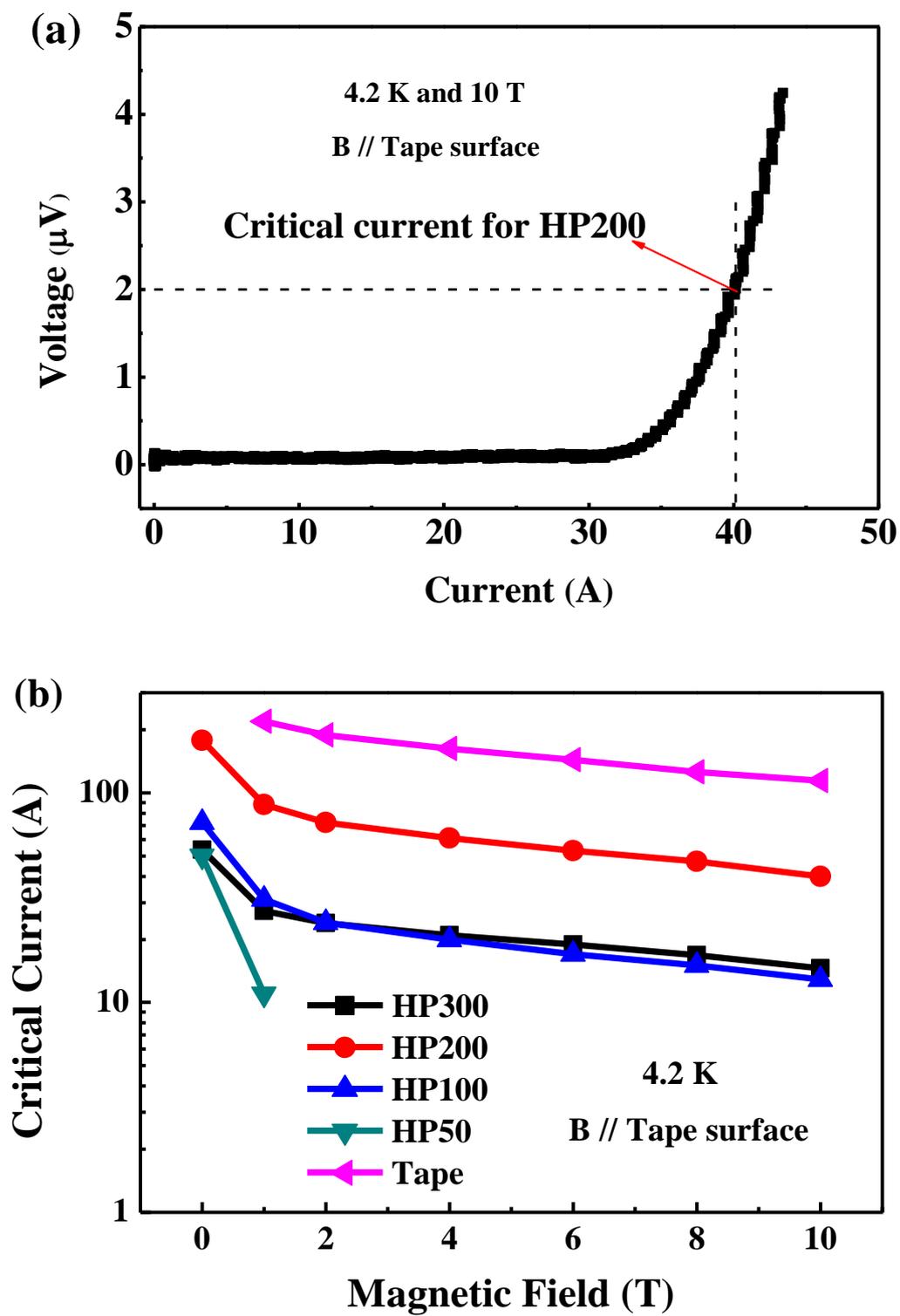

Figure 4. Zhu et al

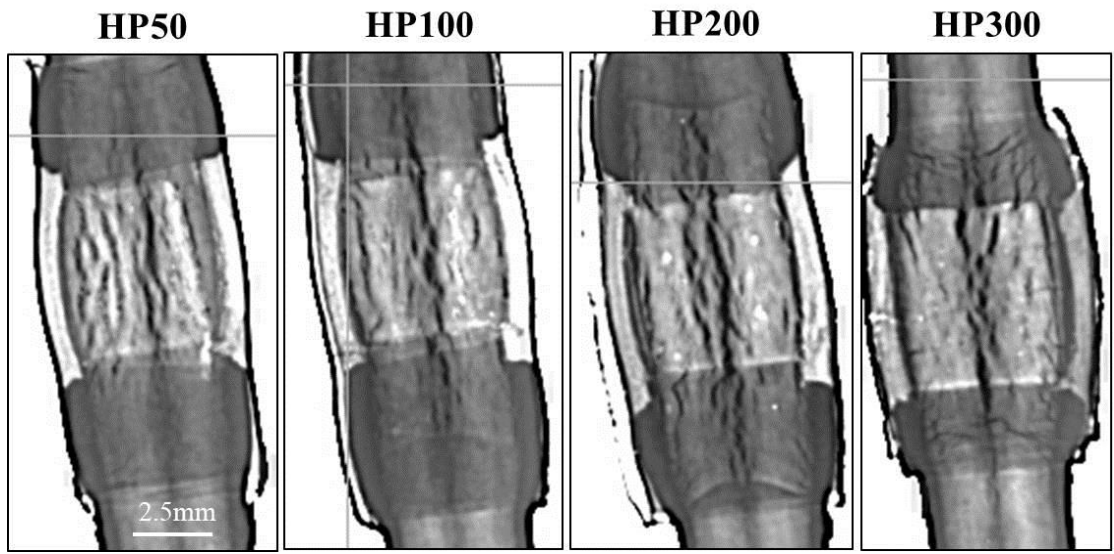

Figure 5. Zhu et al

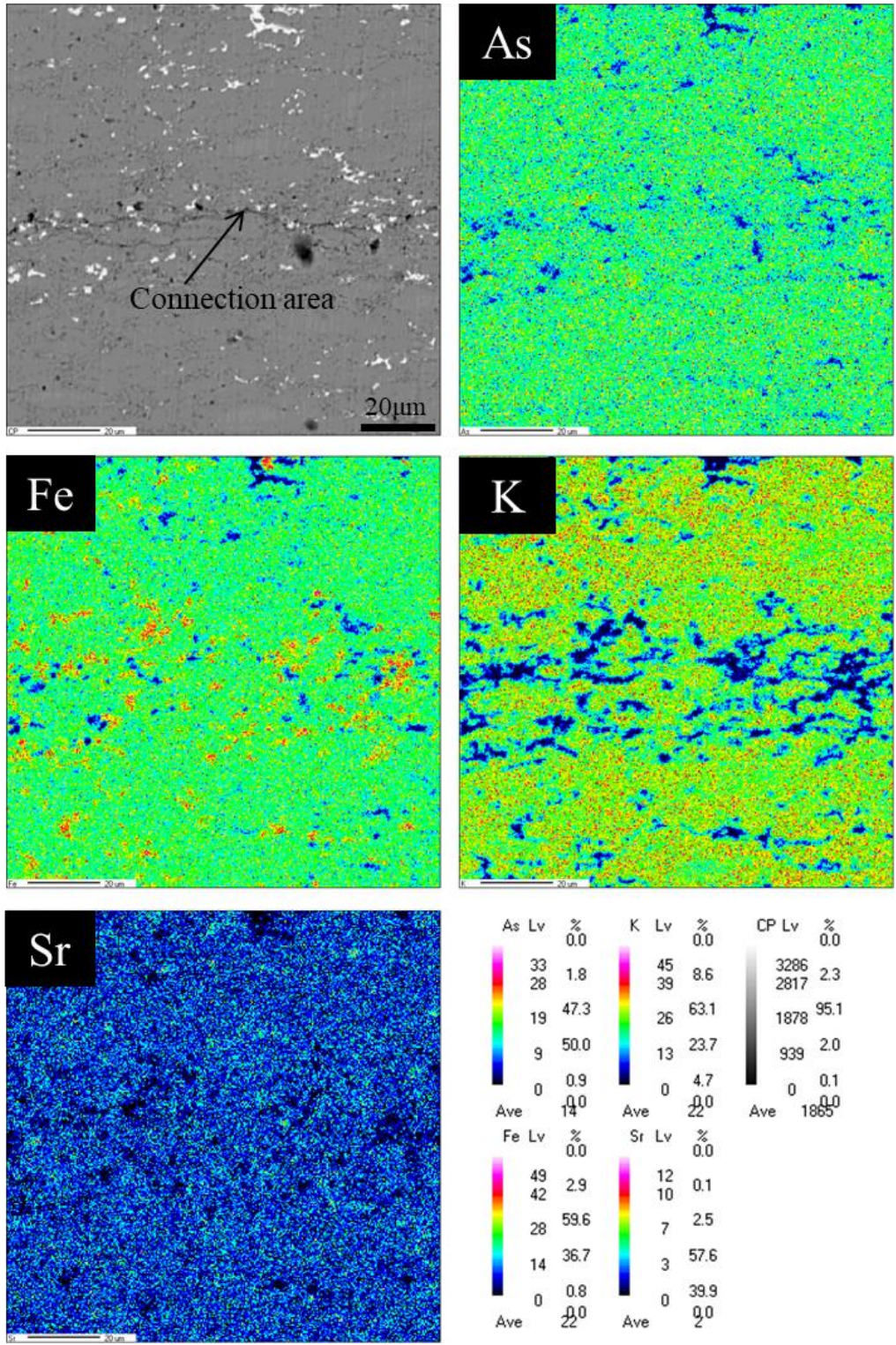

Figure 6. Zhu et al

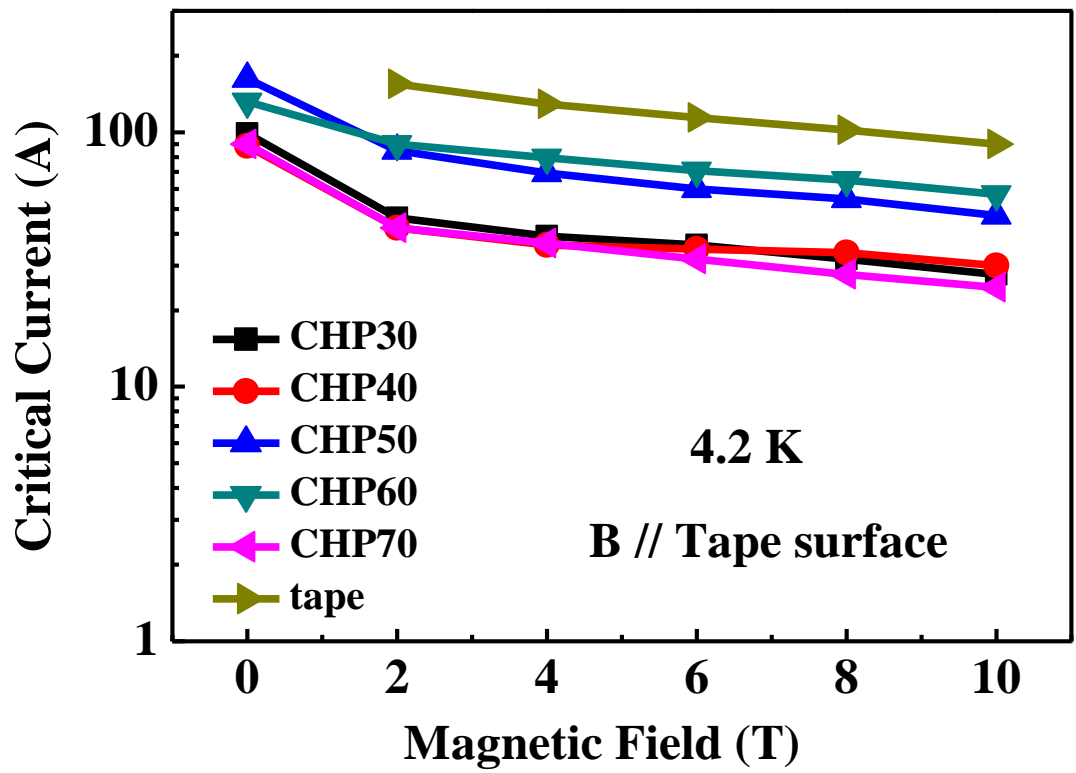

Figure 7. Zhu et al

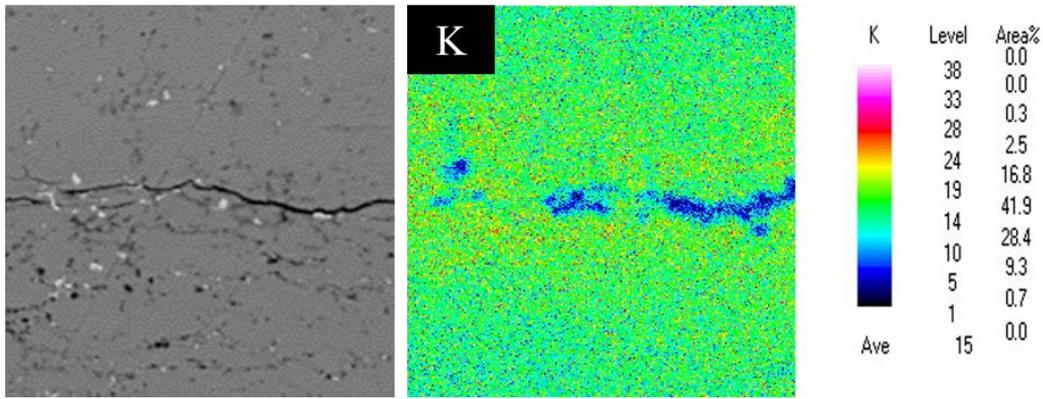

Figure 8. Zhu et al

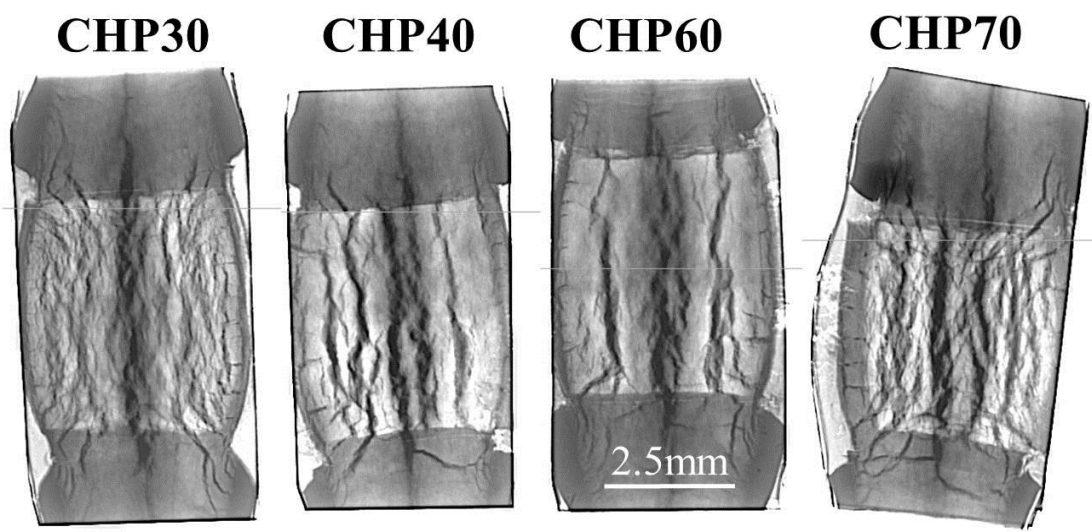

Figure 9. Zhu et al